\documentclass[twocolumn,twoside,slac]{revtex4}
\usepackage{graphicx}
\usepackage{fancyhdr}
\begin{document}

\title
{
GridMonitor: 
Integration of Large Scale Facility Fabric Monitoring with Meta Data Service in Grid Environment\thanks{This work is supported in part by the U.S. Department of Energy and the National Science Foundation.}
}
\author{Rich Baker}
\affiliation{ 
	RHIC/USATLAS Computing Facility \\
        Department of Physics \\
        Brookhaven National Laboratory\\
        Upton, NY 11973, USA}
\author{Dantong Yu} 
\affiliation{ 
	RHIC/USATLAS Computing Facility \\
        Department of Physics \\
        Brookhaven National Laboratory\\
        Upton, NY 11973, USA}
\author{Jason Smith} 
\affiliation{ 
	RHIC/USATLAS Computing Facility \\
        Department of Physics \\
        Brookhaven National Laboratory\\
        Upton, NY 11973, USA}
\author{Anthony Chan}
\affiliation{ 
	RHIC/USATLAS Computing Facility \\
        Department of Physics \\
        Brookhaven National Laboratory\\
        Upton, NY 11973, USA}
\author{Kaushik De}
\affiliation{ 
Department of Physics \\
The University of Texas at Arlington \\
Arlington, TX 76019-0059, USA}

\author{Patrick McGuigan}
\affiliation{ 
Department of Physics \\
The University of Texas at Arlington \\
Arlington, TX 76019-0059, USA}

\begin{abstract}
Grid computing consists of the coordinated use of large sets
of diverse, geographically distributed resources for high
performance computation. Effective monitoring of these
computing resources is extremely 
important to allow efficient use on the Grid.
The large number of heterogeneous computing entities available in Grids
makes the task challenging.  In this work, we describe
a Grid monitoring system, called GridMonitor, that captures and makes 
available the most important information from a large computing facility.
The Grid monitoring system consists of four tiers: local monitoring, 
archiving, publishing and harnessing. This architecture was applied 
on a large scale linux farm and network infrastructure. It can be 
used by many higher-level Grid services including scheduling services 
and resource brokering.

\end{abstract}

\maketitle

\thispagestyle{fancy}

\section{\label{Introduction-ref}Introduction}

Grid computing consists of large sets of diverse, geographically 
distributed resources that are collected into a virtual computer 
for high performance computation. The success of the Grid depends 
greatly on efficient utilization of these resource.  The Particle Physics
Data Grid (PPDG) is an example  
user of the data Grid. PPDG is a collaboration of computer scientists 
in distributed computing and Grid technology, and physicists who work 
on the major high-energy and nuclear physics experiments. These experiments 
include ATLAS, STAR, CMS, D0, and Babar. There are many 
computing resources involved in PPDG physics experiments. 
For example, the computing resources at Brookhaven National Laboratory (BNL)
includes 1100 dual processor PCs which come from six different
vendors. The Linux farms at the BNL RHIC/USATLAS provide 
3.115 TFlops of computation power.  The storage system provides 
140 Tera-Bytes of disk space and 1.2 Peta-Bytes of robotic tape 
storage space. 
The diversity of these computing resources and their large number 
of users make the Grid environment vulnerable to faults and excessive 
loads.  This seriously affects the utilization of Grid resources. 
Therefore, it is crucial to get knowledge about the status of all 
types of computing resources and services to enhance the performance and 
avoid faults.

Here we give an example to illustrate how a Grid application relies on 
Grid information service.  A job scheduler needs information about 
available CPU resources in order to plan the efficient execution of tasks.  
A computing  farm consists of a set of many hosts available 
for scheduling via Grid resource management protocols.  
If required by the exact nature 
of the interrelationship between the farm monitor and the job scheduler,
the hosts at a given site may be broken down into multiple clusters that 
consist of homogeneous nodes, such that the local job manager can 
assume that any queued job can be run on any available node within 
the computing farm. Grid information service should provide the system 
status about each cluster, i.e. cluster configuration, associated
storage system, and so on.  

Many applications, fault detection, performance analysis, 
performance tuning, prediction, and schedule  need information 
about the Grid environment. 
Good methods need to be designed to monitor resource usage, get 
the performance information and detect the potential failures. 
Due to the complexity of the Grid, implementing  a monitoring system 
for such a large scale computing resource is not a trivial task. 
The targets to be monitored in Grid resource include CPU usage, 
disk usage, and network performance of Grid nodes. The ability 
to monitor and manage distributed computing components is critical 
for enabling high-performance distributed computing. Monitoring data 
is needed to determine the source of performance problems and to tune 
the system and application for better performance. Fault detecting 
and recovery mechanisms need monitoring data to determine where the 
problem is, what is the problem and why it happens. 
A performance prediction service might use monitoring 
data as inputs for a prediction model, which would in turn be used 
by a scheduler to determine which resources to use. As more people 
use the Grid, more instrumentation needs will be discovered, 
and more facility status needs to be monitored. 
Many researchers are focused on monitoring computer facility 
in a relatively small scale. The proposed systems are 
Autopilot \cite{RVSR98}, 
Network Weather services \cite{WSH99}, Netlogger \cite{Netlogger97}, 
Grid Monitoring Architecture (GMA) \cite{TCGLT2001} and 
Grid Information Service (MDS) \cite{CFFK2001}.  

Due to the diversity of the computing resource 
and applications in Grid computing, existing monitoring 
architectures can not monitor all of the computing resources 
belonging to the Grid. When the size of a computing facility
grows, the existing monitoring strategy will significantly 
increase the system overhead. The dynamic characteristics 
of the Grid allows the computing resources to participate
and withdraw from the resource pool constantly. Only 
a few existing monitoring systems address this 
characteristic. 
In this work, we present a Grid monitoring system which is adaptive 
to the Grid environment. It includes: 
\begin{itemize}
\item Local monitoring: The local monitoring system 
 monitors the facility which consists of computing, 
 storage and network resources. The monitored information 
 will be provided to different types of application with 
 different requirements. 
\item Grid monitoring: it uses MDS 
 \cite{CFFK2001} to publish the selected monitoring 
 information into the Grid system. 
\end{itemize}
 
 The proposed architecture can separate the facility 
 monitoring from the Grid environment. 
 By using the MDS, it provides
 a well-designed interface between the Grid and the facility. 
 It can provide monitoring information for different
 Grid applications as long they use the Grid information protocols. 
 When new hardware is added to the
 local facility, the local monitoring infrastructure can easily 
 add the new software to monitor the system.  The change of  
 hardware and monitoring tools can be hidden from
 the Grid computing environment. 

The rest of this paper is organized as follows: 
%Section \ref{section-term} introduces the definitions of Grid information service. 
Section \ref{section-system} provide the architecture of 
Grid Monitoring. 
Section \ref{section-relate}
summarizes recent work on Grid monitoring and Grid information 
service. Section \ref{section-conclusion} summarizes our conclusions 
and the scope of future work.

%\section{\label{section-term}Terminology}
%\input{define.tex}
\section{\label{section-system}Grid Monitoring Architecture}

In this section, we specify the system requirements of  
Grid monitoring, provide the Grid monitoring infrastructure, 
and describe the design of each component in the system.  
\subsection{System Requirements}
Due to the complexity and dynamics of the Grid computing model, 
the monitoring toolkits built on top of this computing model 
are also complex. To build an efficient and effective monitoring 
model, the designers and developers need to keep the following 
requirements in mind.
\begin{itemize}
\item
The Monitoring toolkits can make use of existing monitoring tools. 
The overhead for incorporating a monitor tool should be minimum. 
The Grid monitoring system should make use of existing facility
infrastructures with well designed API, not force its own. 
\item
Scalability: the system for monitoring and fault management 
should be scalable. The number of Grid nodes will increase 
every year in order to satisfy the growing computing requirement of HENP. 
The monitoring system should be scalable for the expanding Grid system.
\item
Flexibility: the system for monitoring should be flexible because 
the target to be monitored and the Grid architecture are likely 
to change over time. 
\item
Extensibility and Modularity should be implemented, which allows 
users to include those components easily that they wish to use.
All Communication flows should not flow through a single central 
component. Having a single, centralized repository for dynamic 
data causes two performance problems. The centralized repository 
for information represents a single-point-failure for the entire system. 
The centralized server can create a performance bottleneck.
\item
Non-intrusiveness: the Grid monitoring system should incur as small 
a system overhead as possible. It should not disrupt the normal 
running of the monitored system. This is extremely important 
if a large number of target systems are monitored.
\item
Security: typically, an organization defines policies controlling 
who can access information about their resource. The monitoring 
system must comply with these policies.
\item
Ability of logging: Some important data should be archived. 
\item Inter-operability: Different monitoring systems could obtain and 
 share  each other's monitoring information to avoid the functionality 
 overlapping.  
\end{itemize}
\subsection{A Grid Monitoring Architecture Based on MDS}
\begin{figure*}
\centerline{
 \scalebox{0.48}
 {\includegraphics
{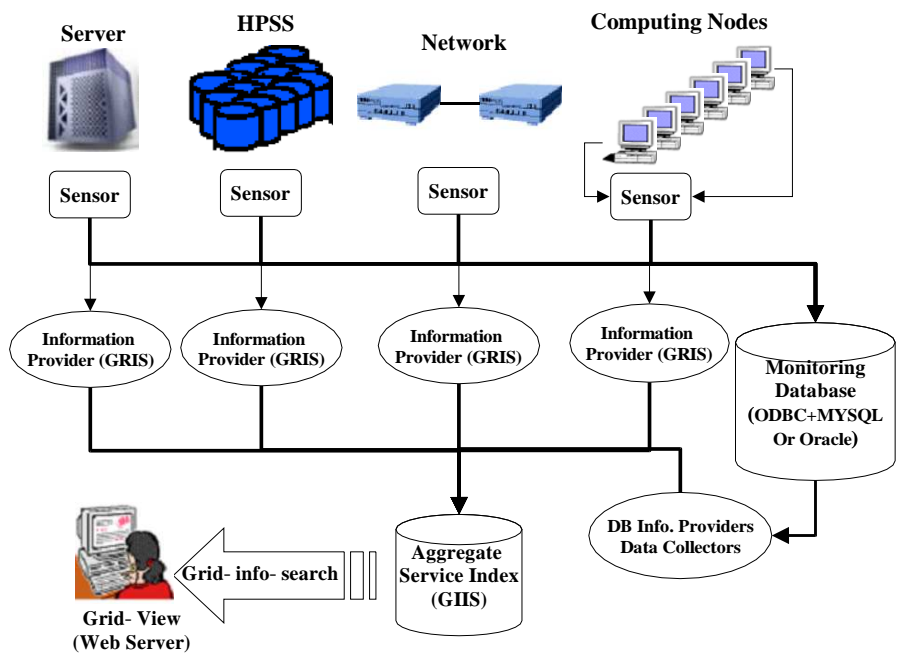}}
}
\caption { Grid Monitoring Framework
         \label{figure-architecture}}
\end{figure*}

Grid monitoring infrastructure has a four-tiered structure: 
Sensors, Archive System, Information Providers and Grid Information
Browser. Figure \ref{figure-architecture} shows how monitoring 
information travels through the four-tiered structure and reaches
end users.  
\begin{itemize}
\item Sensors probe the target systems and obtain related 
statistics. 
\item The data importer of the archive system fetches the statistics 
from sensors and  stores them into the database. 
\item The Grid information provider retrieves information from the database, 
 processes it according to application requirements and returns the results to 
 Grid systems.  
\item The remote web server issues Grid commands to fetch monitoring 
information from the Grid information service.
\end{itemize}

\subsubsection{Sensor}
A sensor can measure the characteristics of a target system.
It generates a time stamped performance statistics. A simple sensor 
example would typically
execute one of the UNIX utilities, such as top, ps, ping, iperf, ndd, or
read directly from system files, such as /proc/* to extract
sensor-specific measurements. 
Sensors are used to monitor CPU usage, memory usage and network traffic. 
Some sensors can monitor and capture abnormal system status. 
We define the type of measurement generated by a sensor, a ``metric''.
As shown in Figure \ref{figure-architecture}, four types of 
computer systems are monitored: file service, high performance storage 
system (HPSS), network equipment and computing nodes.  
Each sensor relies on a set of standard APIs and protocols to 
publish the sensor data. The design of the API and protocol is beyond  
the scope of this paper.

\subsubsection{Archive System}
The Archival system is used to hold historical data that can be used 
for predication and analysis. There are two components in archival 
system: data importer and telemetry database. An importer fetches 
the monitoring data from sensors via standard API, and saves the 
received data into the telemetry database. The database consists mostly 
of telemetry 
gathered by different sensors of different metrics. Some databases 
might include derived parameters, statistics, and any other data 
elements required by the Grid application.  A telemetry database 
can be any relational database, such as Oracle, MySQL, PostgreSQL, 
it can even be a set of flat files. A telemetry database also acts 
as a server 
to answer all types of telemetry queries. Therefore it needs to 
support flexible, complicate query operations and powerful query 
language. SQL is a perfect candidate. But the SQL's for different 
relational databases are different from each other, and  they do not 
support interoperability. To hide the details 
of the underlying databases and provide the user a uniform SQL interface, 
we use ODBC \footnote {http://www.odbc.org} to wrap the telemetry 
databases. This implementation allows greater flexibility in the  
archiving system. 

\subsubsection{Information Provider}
The information provider provides detailed, dynamic statistics about 
instrumentation to Grid monitoring service, MDS. It is managed and
 controlled by MDS. 
The information provider either invokes and stops a set of sensors to 
do active probing, or interacts with running sensors to obtain the 
current status of resource.  An information provider can also query 
the database to get historical information. We implemented our customized
information provider to fetch information from the telemetry database, 
process the information if necessary, and return the necessary
information to the MDS which invokes the information provider. 
  
\subsubsection{Grid Information Browser}
GridView\footnote{http://heppc1.uta.edu/atlas/gridview} 
was developed at the University of Texas at Arlington 
(UTA) to monitor the U.S. ATLAS Grid.  It was the first application 
software developed for the U.S. ATLAS Grid Testbed, released in March, 2001, 
as a demonstration of the Globus 1.1.3 toolkit.  The original text version of GridView provides a 
snapshot of dynamic parameters like cpu load, up time, and idle time for 
all Testbed gatekeepers through a web page, updated periodically. 
MDS information from GRIS/GIIS servers is available through linked pages.
In addition, a MySQL server is used to store archived monitoring information. 
This historical information is also presented through GridView.
A java applet version of GridView is also available.  This applet version
presents a hierarchical display of grid services through a graphical
map-centric view.  As prototype grids become larger and offer more 
services it becomes desirable to have a quick and easy method for 
determining which sites have less than a complete set of operational 
services, along with detailed error messages for services that are failing.  
GridView fulfills this need by performing tests of the Globus Toolkit 
based services at grid computing resources and presenting the 
results via an applet that provides different views of the state of the testbed.

GridView is comprised of two different subsystems, a data collection 
daemon and a Java applet for visualization.  The data collection process 
periodically contacts remote computing systems to ascertain the operational 
status of the three services offered by the Globus Toolkit.  During the testing 
at the remote system, a transcript is maintained of the tests 
performed, the status of the tests and any generated error messages that 
indicate faulty services.  Also saved during testing is the information 
provided by the Globus Monitoring and Discovery Service (MDS) for 
both Grid Information Index Services (GIIS) and Grid Resource 
Information Services (GRIS).  At the completion of a testing cycle, the 
data collection daemon publishes this information to an HTTP server that 
provides the applet to users.

\begin{figure*}
\centerline{
 \scalebox{0.67}
 {\includegraphics
{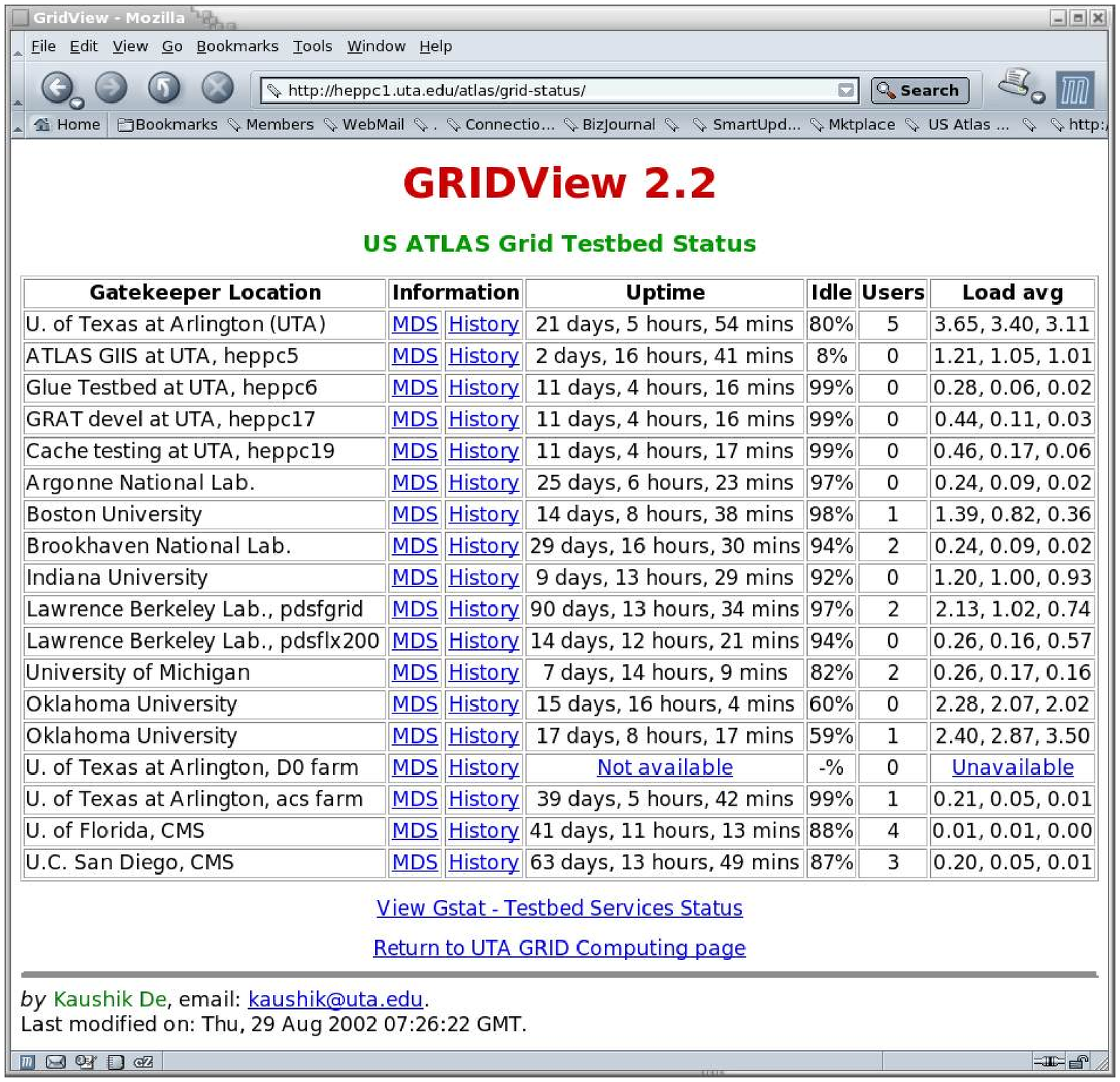}}
}
\caption {Grid View Interface (text version) 
         \label{figure-gridview}} 
\end{figure*}

\begin{figure*}
\centerline{
 \scalebox{0.67}
 {\includegraphics
{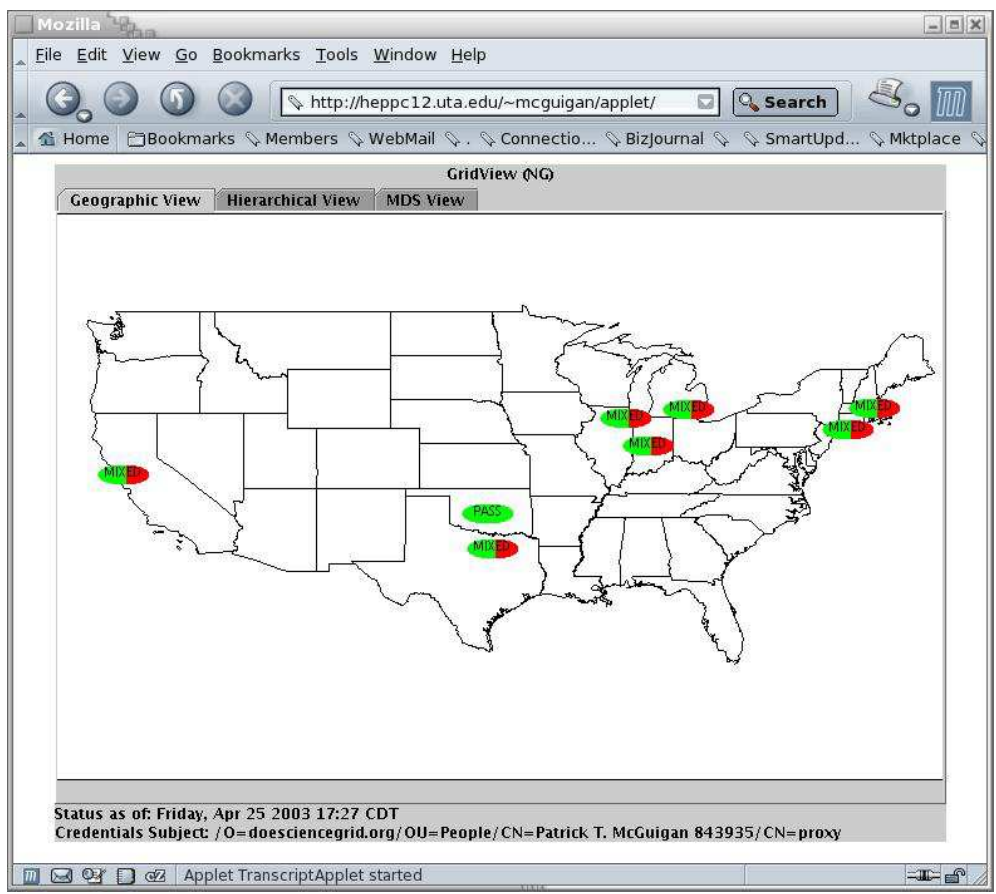}}
}
\caption {Grid View Interface (Map View) 
         \label{figure-gridview-map}} 
\end{figure*}

\begin{figure*}
\centerline{
 \scalebox{0.67}
 {\includegraphics
{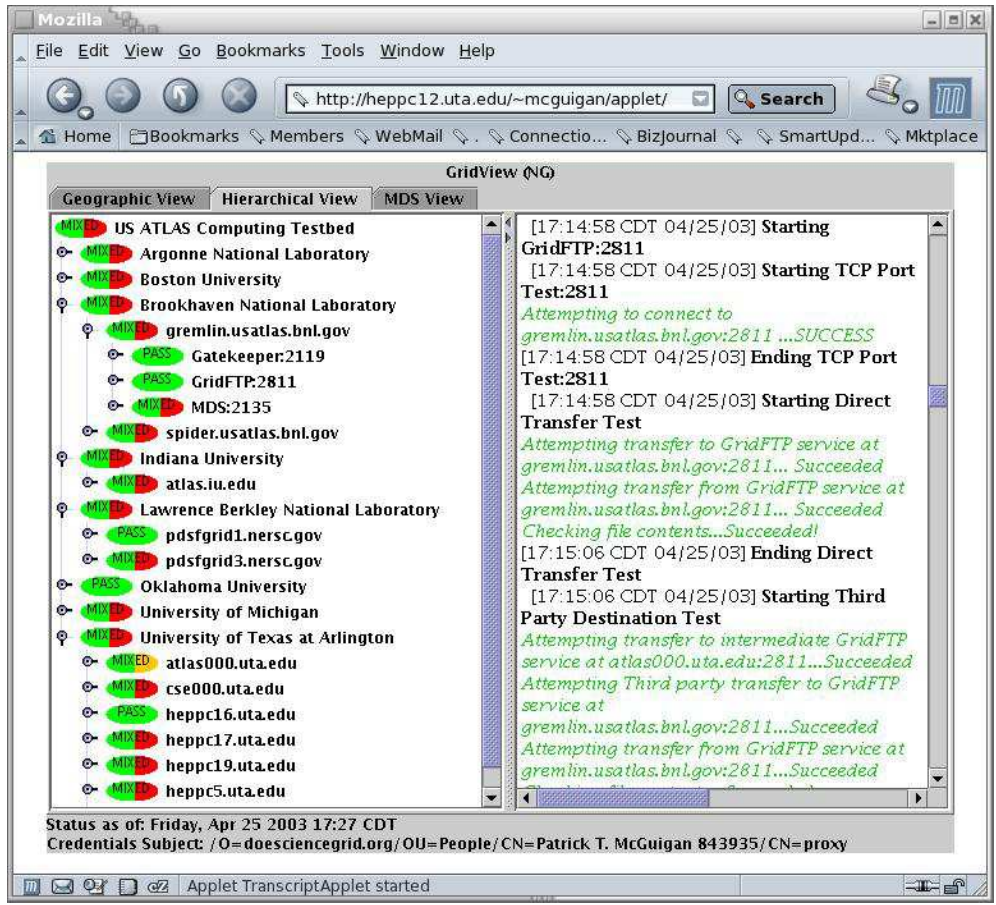}}
}
\caption {Grid View Interface for the Hierarchical Service Status
         \label{figure-gridview-hie}} 
\end{figure*}

\begin{figure*}
\centerline{
 \scalebox{0.67}
 {\includegraphics
{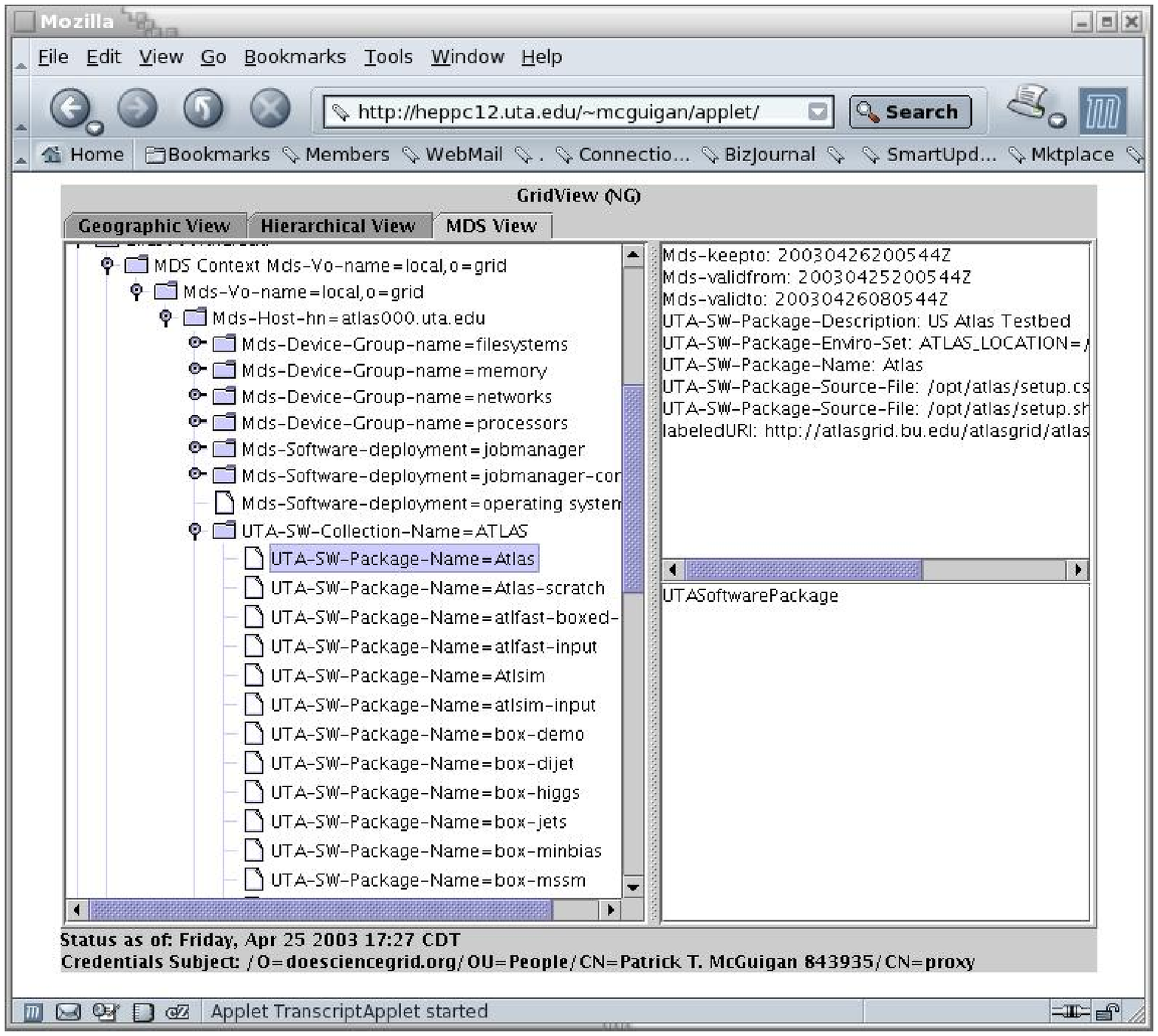}}
}
\caption {Grid View Interface for MDS Information 
         \label{figure-gridview-mds}} 
\end{figure*}

The visualization applet uses the information recorded by the data collection 
daemon and presents it in three differing views via an HTTP server.  
Figure \ref{figure-gridview-map} presents a geographical representation 
where top-level computing sites are shown on a map of the continental United States.  
A status icon for each site shows the combined status of all computers tested at the 
site.  This view provides a quick snapshot of the overall status of the U.S. 
ATLAS computing testbed.
A hierarchical view of the data shows users the status information for every 
test and test sequence along with the related transcripts associated with 
the tests.  By following the color-coded status icons, users can quickly 
determine which tests failed at which sites.  Clicking on a particular 
test or test sequence will automatically bring the associated test transcript into view, as shown Figure \ref{figure-gridview-hie}.
Finally, the users can inspect the contents of the MDS services offered in 
the testbed, as shown in Figure \ref{figure-gridview-mds}.  
This is a graphical and hierarchical view of the data retrieved 
during testing.  The main goal of this screen is to allow users to 
view static MDS entries as well as the exact values of dynamic 
entries that have violated the LDAP consistency tests.
Figure \ref{figure-gridview} shows 
the status of USATLAS Grid testbed as viewed through the GridView 
text interface. 

\subsubsection{Desired Features of the System Architecture}
The proposed architecture can separate the facility
monitoring from the Grid environment.  
The changes in  hardware and monitoring tools can be hidden from
the Grid computing environment. 
When new hardware is added in the
local facility, the local monitoring infrastructure can easily
add the new hardware to the monitoring system.  
This monitoring system simplifies the design of new sensors: 
new sensors can be plugged into the monitoring architecture with 
minimum effort. 
%the sensors only need to passively answer any 
%inquiry or broadcast at a certain interval.
The sensors do not need to know who wants to subscribe to this metric
and the number of the subscribers. The subscribers (consumers) 
can also be simplified because they just need to tell information 
provider what metric they are interested in. It is up to the information 
provider to deliver the specified metrics to the subscribers. 
The MDS information provider
provides a well-designed interface between the Grid and computing facilities. 
It can provide monitoring information for different
Grid applications as long they use protocols provided
by the MDS \cite{CFFK2001}. 
By distributing and replicating the telemetry databases and MDS 
servers in different locations, we can avoid the problems caused by a 
single centralized server.

\begin{itemize}
\item The MDS provides the cache copy of the lastest value from the 
MySQL database.
\item Non-intrusiveness: 
Sensors and local monitoring tools put less than 1 Percent CPU Load
on the entire system. 
The information provider can prevent users from directly accessing to 
the database server, protect the sensitive information in the database 
effectively.  
\item Scalability: 
1100 linux nodes and the network connectivity of eight USATLAS testbeds 
 are monitored by our system without adding too much load on the target systems.
\item Flexibility:
Independent of Sensors. Many sensors can be easily plugged in as long 
as they have a well defined protocol and API. 
Another advantage is that the archival system is independent to the 
underlying database.
\end{itemize}

\section{\label{section-relate}Related Works}
{\bf Ganglia}: Ganglia \cite {ganglia} is a scalable distributed 
monitoring system 
designed for high performance computing systems such as large 
localized clusters and even widely distributed Grids.  
It relies on a multicast-based listen/announce protocol to monitor 
the state within clusters and uses point-to-point connections among
representative cluster nodes to federate clusters into a Grid 
and aggregate their states. Ganglia has the advantages of low 
per-node overhead, high concurrency and robustness.  

Ganglia has been deployed at BNL on the RHIC and Atlas clusters to
successfully monitor over 1000 nodes, organized into 10 separate
clusters based on experiment.  One collection node has been setup
to gather all of the data from these clusters and archive it locally to
be displayed by ganglia's web front end.  The data is made available to
each experiment for their own monitoring and job scheduling needs and is
also published through Globus MDS.

{\bf Network Weather Service (NWS)}: The goal of the Network Weather Service 
is to provide accurate forecasts of dynamically changing performance 
characteristics from a distributed set of meta-computing resources. 
It can produce short-term performance forecast based on historical 
performance measurement. The Network Weather Service attempts to 
use both existing performance monitoring utilities and its own active 
sensors to make use of resource, probe its own usage and measure the 
performance. 
It can measure the fraction of 
CPU time available for new processes, TCP connection time, end-to-end 
TCP network latency, and end-to-end TCP network bandwidth. It has NWS 
sensors, CPU sensors, network sensors. It also has predictors that 
forecast the system performance. NWS was widely adopted by many Grid 
communities. Therefore, we will incorporate NWS in our 
Grid monitoring toolkits. We can pull out the sensor modules and prediction 
modules and put them into our monitor architecture. We also need to design 
an interface that can bridge the communication between the sensors and 
the telemetry database. 

{\noindent {\bf Simple Network Management Protocol (SNMP)}}:
Since SNMP was developed in 1988, the Simple Network Management 
Protocol has become the standard for inter-network management. 
Because it is a simple solution, requiring little code to 
implement, we can easily build SNMP sensors for our monitoring 
architecture. SNMP is extensible, allowing us to easily add 
network management functions to the monitoring system. SNMP 
also separates the management architecture from the architecture 
of the hardware devices, which broadens the arena of our monitoring 
architecture. SNMP is widely available today and has extensive 
support from academic, commercial vendors and research institutes. 
Therefore, SNMP based tools are widely used for network monitoring 
and management. SNMP based tools and sensors should be evaluated 
for our Grid monitoring system.

{\noindent {\bf Monitoring and Discovery Service (MDS)}}:
MDS provides the Grid information in Globus \cite{gridbook}
and  OGSA \cite{OGSA}.  The MDS stores the information 
collected by its information providers in a cache. These 
information providers are run periodically to update the information 
about the hosts, networks, memory usage, disk storage and software
available on the system and batch queue status. MDS is
designed to monitor large number of  entities and help users to 
discover and keep track of these resources. It supports a registration 
protocol which allows individual entities and their information providers
to join and leave MDS dynamically. The monitoring 
infrastructure is organized hierarchically, built on top of 
the LDAP server (light weight directory access protocol \cite{ldap-1, ldap-2}). 
MDS provides LDAP compatible client tools to access the MDS server.
Due to its LDAP-based implementation, MDS is not designed to handle 
highly volatile monitoring data. 

{\noindent {\bf Grid Monitoring Architecture (GMA)}}:
 The Grid Monitoring Architecture consists of three components: 
 directory service, producer and consumer. Producers publish their 
 existence, description and type of performance data to the directory 
 service. Consumers query the directory service and subscribe to 
 the selected producer. The time-stamped performance data, 
 called events, are directly sent from the producers to consumers 
 based on subscription entries stored at the directory service. 
 Grid Monitoring Architecture supports both a streaming publish/subscribe
 model, and query/response model. Compared with MDS, the GMA supports the 
 highly dynamic monitoring data. The data stream continuously flows 
 from producers to consumers until the subscription becomes invalid.

\section{\label{section-conclusion}Conclusion}

Grid computing benefits from a scalable monitoring system. 
GridMonitor is a promising candidate for this role. It can be 
used to monitor  several thousand computers, geographically 
distributed among several computing centers.  
It naturally integrates large scale 
fabrication monitoring into the grid system. 
The initial prototype was deployed at Brookhaven National Laboratory.  
Open challenges include performance, availability
of crucial system status information, robustness  and scalability.  

\begin{acknowledgments}
The authors wish to thank RHIC/USATLAS computing facility group
for their valuable comments and discusses for this work.
This work is supported by grants from the U.S. Department
of Energy and the National Science Foundation.
\end{acknowledgments}

%\begin{thebibliography}{99} % Use for 10-99 references

\end{document}